\documentstyle[aps,epsf]{revtex}

\begin{document}

\title{Form and width of spectral line
of Josephson Flux-Flow oscillator}
\author{Andrey L. Pankratov}
\address{Institute for Physics of Microstructures of
RAS, Nizhny Novgorod, RUSSIA. \\
E-mail: alp@ipm.sci-nnov.ru}

\maketitle

\begin{abstract}
The behavior of a Josephson flux-flow oscillator in the presence
of both bias current and magnetic field fluctuations has
been studied. To derive the equation for slow phase dynamics in
the limit of small noise intensity the Poincare method has been
used. Both the form of spectral line and the linewidth of the
flux-flow oscillator have been derived exactly on the basis of
technique presented in the book of Malakhov \footnote{This paper
is dedicated to memory of my teacher, The Honored Scientist of
Russia, Russian State Prize Winner, Prof. Askold N. Malakhov
(5.12.1926-7.11.2000).} \cite{mal}, known limiting cases are
considered, limits of their applicability are discussed and
appearance of excess noise is explained. Good coincidence of
theoretical description with experimental results has been
demonstrated.

\noindent PACS number(s): $74.50.+r, 74.40.+k$
\end{abstract}



\section{Introduction}

Long Josephson oscillators operating in the flux-flow regime
\cite{nei} are presently considered as possible devices for
applications in superconducting millimeter-wave electronics
\cite{ksh}. In comparison to single fluxon oscillators they have
higher output power, wider bandwidth, and easier tunability, but
they have a wider linewidth \cite{linewidth} of the emitted
radiation from the junction. Recent measurements by Koshelets et
al. \cite{koshelets,kosh2} have indeed shown a linewidth for a
Josephson flux-flow oscillator which is of about one order of
magnitude wider than the one derived for a short (lumped)
Josephson junction \cite{RS,barone,lik}. This last property is
quite undesirable if one wishes to use such devices, for
example, as local oscillators in radioastronomy receivers
\cite{ksh}. For concrete applications it is important to get a
model which adequately describes the linewidth of the flux-flow
oscillator. With such a model one can hope to control the
phenomenon of linewidth broadening by properly choosing the
design parameters of the device. A first attempt in this
direction was performed by Golubov et al. and Ustinov et al.
\cite{gmu96,ukh96} in terms of a particle model for the train of
fluxons moving in the junction. However, the authors of
\cite{gmu96} calculated the variance of frequency fluctuations
(that is quantity difficult to measure) but not a linewidth.
Another attempt to derive the linewidth was performed in
\cite{bk97}, but the results obtained are restricted by the
consideration of "particle-like" picture of fluxon motion in an
infinite junction and the linewidth is expressed in quantities
that are not easily accessible from experiment in a flux-flow regime,
e.g., average interval between fluxons. Besides, magnetic field
fluctuations and parametric effects that may lead to additional
broadening of the linewidth, were not considered in those papers.
Recently, importance of accounting magnetic field fluctuations
has been discussed in \cite{koshisec}.

The task of deriving of the linewidth of FFO may be decomposed
into two parts: one is more experimental and another one
is more methodological.

One of the difficulties of the considered
problem is absence of understanding of nature of noises of FFO.
It is clear, that there are several different noise sources
influencing the FFO: natural wideband noises (such as thermal
and shot noises), technical narrowband noises and possibly
flicker noise. And all these noise sources affect the FFO
via fluctuations of both bias current and magnetic field
(control line current).
Considering present FFO designs \cite{kosh3}, one can guess, that
there are some noise components in bias and control line currents
that are correlated and some that are uncorrelated and lots of
detailed experimental study should be performed to understand
nature of this complicated mixture of fluctuations.

On the other hand, if we know the parameters of noise
(say, parameters of both natural and technical
fluctuations), our task is to obtain the linewidth and
the form of spectral line of FFO, and on the basis of
the obtained characteristics to predict how to improve
the noise properties of the oscillator. And here we
would like to consider this "methodological" part
of the analysis of noise properties of FFO.

The aim of the present paper is to give strictly mathematical
derivation of the required fluctuational equation for slow
component of the phase that can be done for the most important
case of small noise intensity, and present exact derivation of
the linewidth and the form of spectral line on the basis of
methods described in the book by Malakhov \cite{mal}. We will
assume a certain model of noise sources of FFO bias and control
line current fluctuations, as well as we will consider the
parametric effect of higher harmonics leading to additional
broadening of the spectral line.

\section{Basic equations}

The electrodynamics of a long Josephson junction in the presence
of magnetic field is described by the perturbed sine-Gordon
equation
\begin{equation}
\frac{\partial^2\phi}{\partial t^2}+\alpha\frac{\partial\phi}
{\partial t}-\frac{\partial^2\phi}{\partial x^2}=\eta-\sin (\phi)
\label{PSGE}
\end{equation}
subject to the boundary conditions
\begin{equation}
\frac{\partial\phi(0,t)}{\partial x}=
\frac{\partial\phi(L,t)}{\partial x}=\Gamma.  \label{bc}
\end{equation}
In this equation space and time have been normalized to the
Josephson penetration length $\lambda _{J}$ and to the inverse
plasma frequency $\omega_{p}^{-1}$, respectively, $\alpha$ is the
loss parameter, $\eta$ is the normalized dc bias current density
and $\Gamma$ is the normalized magnetic field. In accordance with
RSJ model \cite{lik} one takes the loss parameter
$\alpha=\frac{\omega_{p}}{\omega_{c}}$, where
$\omega_p=\sqrt{2eI_c/\hbar C}$, $\omega_{c}=2eI_cR_{N}/\hbar$,
$C$ is the capacitance, $R_N$ is the normal state resistance
($R_N=V/I_{qp}$, $V$ being voltage and $I_{qp}$ -- the
quasiparticle component of the current), $I_c$ is the critical
current, $\eta=J/J_c$ ($I=\int\limits_0^l J(x)dx$,
$I_c=\int\limits_0^l J_c(x)dx$, $I$ is the bias current), $l$ is
dimensional length of the junction, $L=l/\lambda_J$.

In general, both bias current $\eta$ and magnetic field $\Gamma$
(control line current) may fluctuate: $\eta=\eta_0+\eta_F(x,t)$,
$\Gamma=\Gamma_0+\Gamma_F(x,t)$ and usually these fluctuations are
supposed to be wideband noises and are small:
for $\eta_0\ne 0$ and $\Gamma_0\ne 0$ variances of
$\eta_F(x,t)/\eta_0$, $\Gamma_F(x,t)/\Gamma_0$ are much smaller
than unity.
Therefore, we will consider the noise sources $\eta_F(x,t)$ and
$\Gamma_F(x,t)$ as perturbations that do not affect the
current-voltage characteristic, but lead to nonzero width of the
spectral line. Following \cite{lik}, we suppose that
$\eta_F(x,t)$ is Gaussian noise with zero mean value
$\left<\eta_F(x,t)\right>=0$ and its spectral density is so wide
that $\eta_F(x,t)$ may be treated as white noise with the
correlation function:
\begin{equation}
\left<\eta_F(x,t)\eta_F(x',t')\right>=\frac{2k_B T\omega_p}
{R_N I_c J_c \lambda_J} \delta (x-x^{\prime})\delta (t-t^{\prime}).
\end{equation}
Here and in the following $<>$ denotes ensemble average, $k_{B}$
is the Boltzmann constant and $T$ is the temperature.
In comparison with \cite{RS}, we
consider simple RSJ model for current fluctuations:
usually, at standard working temperature $T=4.2K$,
pair current fluctuations are much smaller than
quasiparticle-current fluctuations and may be neglected:
$I_p=0$. Also we do not take into consideration the shot noise
contribution that may be neglected if the condition $2k_B T\gg eV$
is fulfilled.

The properties of the magnetic field fluctuations
$\Gamma_F(x,t)$ were not studied in the literature. From the
present designs of FFO \cite{koshelets,kosh2,kosh3} one can, however,
make some conclusions about nature of these fluctuations. In the
present layouts the base electrode of the long Josephson junction
is employed as a control line. Therefore, wideband bias current
fluctuations will enter the control line. Moreover, following
recent idea of Koshelets (experimentally confirmed in
\cite{koshisec}), even if the control line is isolated
from the junction, fluctuating bias current may induce magnetic
field, that will affect fluxons. On the other hand, narrowband
technical fluctuations also exist there. So, we can model the
control line fluctuations as follows:
$\Gamma_F(x,t)=\sigma\eta_F(x,t)+\Gamma_I(x,t)+\Gamma_T(x,t)$,
where $\Gamma_I(x,t)$ are internal control line pair-current
fluctuations (we neglect by
$\Gamma_I(x,t)=0$, supposing that they are much smaller than
$\sigma\eta_F(x,t)$) and $\Gamma_T(x,t)$ are
narrowband technical fluctuations.
Since there are many compensation techniques that allow to
significantly eliminate influence of narrowband technical
fluctuations \cite{lik,koshelets,kosh2,koshisec}, we will also
neglect them: $\Gamma_T(x,t)=0$. The question about attenuation
factor $\sigma$ via which bias current fluctuations are converted
into magnetic field fluctuations is not trivial and lot of
theoretical and especially experimental work should be done to
answer this question. It is clear, that $\sigma$ will be
different for different types of FFOs and depends on the junction
geometry and distribution of currents in the base electrode. Let
us suppose, that the value of $\sigma$ is known and later we will
discuss how $\sigma$ may be measured. We note, that the approach
for linewidth calculation \cite{mal} presented below recently was
successfully used for calculation of the linewidth of Cherenkov
FFO \cite{PRB00}. This approach is rather universal and allows to
take into account almost any noise sources, even flicker noise
(for which power spectral density diverges for $\omega\to 0$,
but, nevertheless, calculation of the linewidth may be done in
this case \cite{mal}) and in the present paper we neglect
technical fluctuations only to simplify the analysis.

The flux-flow regime is characterized by excitations which
travel on top of a fast rotating background so that the
effective nonlinearity in the system is drastically reduced
due to fulfilling the following conditions: $\eta/\alpha\gg 1$
and (or) $\Gamma\gg 1$.
In order to derive the linearized equation for slow component of
the phase $\phi(x,t)$ (that is required for obtaining the spectral
characteristics) we will first derive it in the case of zero noise
intensity ($\eta_F(x,t)=\Gamma_F(x,t)=0$) and later will consider
noise as small perturbation.
In papers \cite{cgssv98}, \cite{ss99} linear mode theory and
perturbative analysis around rotating background
($\phi=\phi_0+\psi$, $\psi\ll 1$) have been
used to derive the current-voltage characteristic of FFO.

We will use more general Poincare method: obtain the solution as the
series with respect to the small parameter
$\epsilon=\left(\frac{\alpha}{\eta}\right)^2\ll 1$.
Let us change variables in Eq. (\ref{PSGE}),
$\tau=\frac{\eta}{\alpha}t$, $z=\frac{\eta}{\alpha}x$:
\begin{equation}
\frac{\partial^2\phi}{\partial\tau^2}+
\beta \frac{\partial\phi}{\partial\tau}-
\frac{\partial^2\phi}{\partial z^2}=
\beta-\epsilon\sin (\phi),
\label{SEI}
\end{equation}
where $\beta=\alpha^2/\eta$.

The steady-state solution of this equation we will
find in the form:
$\phi(\tau)=\phi_0(\tau)+\epsilon\phi_1(\tau)+
\epsilon^2\phi_2(\tau)+\dots$
($|\phi_0(\tau)|\gg\epsilon|\phi_1(\tau)|\gg
\epsilon^2|\phi_2(\tau)|\gg\dots$).
Substituting this into Eq. (\ref{SEI}) we will find the zero
order equation:
\begin{equation}
\frac{\partial^2\phi_0}{\partial\tau^2}+
\beta\frac{\partial\phi_0}{\partial\tau}-
\frac{\partial^2\phi_0}{\partial z^2}=\beta.
\label{SE0}
\end{equation}
It is easy to see, that the steady-state solution of
this equation is: $\phi_0(\tau)=\tau+\gamma z=
\frac{\eta}{\alpha}t+\Gamma x$, $\gamma=\alpha\Gamma/\eta$.
To get higher order equations we have to decompose
$\sin(\phi_0(\tau,z)+\epsilon\phi_1(\tau,z)+
\epsilon^2\phi_2(\tau,z)+\dots)$ into Taylor expansion.
From the structure of the considered linear recurrent
equations we know, that the steady-state solution
$\phi_n(\tau,z)$ may be presented in the form:
$\phi_n(\tau,z)=\omega_n \tau+\phi_{np}(\tau,z)$, where
$\phi_{np}(\tau,z)$ is periodic nongrowing component.

Let us now collect together all linearly growing components
$\omega_n\tau$ and we will get: $\sin(\phi(\tau,z))=
\sin(\{\omega_0\tau+\epsilon\omega_1\tau+\epsilon^2\omega_2\tau
+\dots+\gamma z \}+\epsilon\phi_{1p}(\tau)+
\epsilon^2\phi_{2p}(\tau)+\dots)$. Now we can linearize
$\sin(\phi)$ as: $\sin(\phi)\approx \sin(\omega_J\tau+\gamma
z)+\epsilon\phi_{1p}(\tau,z) \cos(\omega_J\tau+\gamma z)+
\epsilon^2\phi_{2p}(\tau,z) \cos(\omega_J\tau+\gamma z)+\dots$,
where
$\omega_J=\omega_0+\epsilon\omega_1+\epsilon^2\omega_2+\dots$ is
the oscillation frequency ($\omega_0=1$), and $\omega_1$,
$\omega_2$,..., $\omega_n$,..., $\phi_{1p}(\tau,z)$,
$\phi_{2p}(\tau,z)$, ..., $\phi_{np}(\tau,z)$, ... are unknown
functions that we want to obtain. Restricting ourselves by
consideration the solution up to the 2-nd order only (in principle
we can do it up to any order, all equations may be solved
recursively), we get the following equations for
$\phi_{1}(\tau,z)$, $\phi_{2}(\tau,z)$:
\begin{equation}
\frac{\partial^2\phi_1}{\partial\tau^2}+\beta
\frac{\partial\phi_1}{\partial\tau}-\frac{\partial^2\phi_1}{\partial z^2}=
-\sin(\omega_J\tau+\gamma z),
\label{SE1}
\end{equation}
\begin{equation}
\frac{\partial^2\phi_2}{\partial\tau^2}
+\beta\frac{\partial\phi_2}{\partial\tau}
-\frac{\partial^2\phi_2}{\partial z^2}=
-\phi_{1p}(\tau)\cos(\omega_J\tau+\gamma z).
\label{SE2}
\end{equation}
It is easy to see from Eq. (\ref{SE1}) that $\omega_1=0$
and substituting the solution in the form
$$\displaystyle{\phi_{1p}(\tau,z)=\sum\limits_{n=0}^{\infty}
\left[ \overline{A}_{1n}\cos(\omega_J\tau)+
\overline{B}_{1n}\sin(\omega_J\tau)\right]
\cos(\overline{k}_n z)}$$ into (\ref{SE1}) one can find $\overline{A}_{1n}$
and $\overline{B}_{1n}$:
\begin{equation}
\overline{A}_{1n}=(2-\delta_{0,n})\frac{\beta\omega_J I_C-
(\overline{k}_{n}^2-\omega_J^2)I_S}
{(\beta\omega_J)^2+(\overline{k}_{n}^2-\omega_J^2)^2},
\label{A}
\end{equation}
\begin{equation}
\overline{B}_{1n}=-(2-\delta_{0,n})\frac{\beta\omega_J I_S+
(\overline{k}_{n}^2-\omega_J^2)I_C}
{(\beta\omega_J)^2+(\overline{k}_{n}^2-\omega_J^2)^2},
\label{B}
\end{equation}
\begin{equation}
I_S=\frac{1}{\overline L}\int_0^{\overline L}
\sin(\gamma z)\cos(\overline{k}_{n} z) dz, \,
I_C=\frac{1}{\overline L}\int_0^{\overline L}
\cos(\gamma z)\cos(\overline{k}_{n} z) dz,
\overline L=\frac{\eta}{\alpha}L.
\label{I}
\end{equation}

Substituting this solution (where $\omega_J=1+
\epsilon^2\omega_2$) into Eq. (\ref{SE2}), we get:
\begin{equation}
\omega_2=-\sum\limits_{n=0}^{\infty}
\frac{2-\delta_{0,n}}{2}\left[\frac{(1+\epsilon^2\omega_2)
[I_S^2+I_C^2]}{(\beta(1+\epsilon^2\omega_2))^2+
[\overline{k}_n^2-(1+\epsilon^2\omega_2)^2]^2}\right].
\label{w2}
\end{equation}

From this equation $\omega_2$ may be found. Analogically to
$\phi_{1p}(\tau)$ one can find $\phi_{2p}(\tau)$.

Combining Eqs. (\ref{SE0})-(\ref{SE2}) together we can get
equation for $\psi(\tau,z)=\phi_0(\tau,z)+\epsilon\phi_1(\tau,z)+
\epsilon^2\phi_2(\tau,z)+\dots+\epsilon^n\phi_n(\tau,z)$ that
is equivalent to the steady-state case of the original equation
(\ref{PSGE}) up to the $n$-th order and
$\displaystyle{\lim_{n\to\infty}\psi(\tau,z)=\phi(\tau,z)}$:
\begin{eqnarray}
\frac{\partial^2\psi}{\partial\tau^2}+\beta
\frac{\partial\psi}{\partial\tau}-
\frac{\partial^2\psi}{\partial z^2}
=\beta-\epsilon\sin(\omega_J\tau+\gamma z)-
\frac{\epsilon^2}{2\overline{L}}\int\limits_0^{\overline{L}}
\sum\limits_{n=0}^{\infty}\left[\overline{A}_{1n} \cos(\gamma z)
-\overline{B}_{1n} \sin(\gamma z)\right]\cos(\overline{k}_n z)dz-
\nonumber \\
-\frac{\epsilon^2}{2}\sum\limits_{n=0}^{\infty}\left[
\overline{A}_{1n} \cos(2\omega_J\tau+\gamma z)
+\overline{B}_{1n} \sin(2\omega_J\tau+\gamma z)\right]
\cos(\overline{k}_n z)-\dots,
\label{SEN}
\end{eqnarray}
where $\omega_J=1+\epsilon\omega_1+\epsilon^2\omega_2+
\dots+\epsilon^n\omega_n$ is supposed to be known.
Here we have presented in the explicit form the term
$\phi_{1p}(\tau)\cos(\omega_J\tau+\gamma z)$ and have
taken into account that only zero mode term with $m=0$
will contribute into linearly growing component of
$\phi_2(\tau)=\sum\limits_{m=0}^{\infty}\overline{A}_{2m}(\tau)
\cos(\overline{k}_m z)$, whereas all contributions due to
$\frac{1}{\overline{L}}\int\limits_0^{\overline{L}}
\sum\limits_{n=0}^{\infty}\left[\overline{A}_{1n} \cos(\gamma z)
-\overline{B}_{1n} \sin(\gamma z)\right]\cos(\overline{k}_n z)
\cos(\overline{k}_m z)dz$, $m\ne 0$, will decay with time.

Now, before introducing noise sources, it is convenient
to change variables back:
\begin{eqnarray}
\frac{\partial^2\psi}{\partial t^2}+\alpha
\frac{\partial\psi}{\partial t}- \frac{\partial^2\psi}{\partial
x^2} =\eta+\alpha\Omega_2-f(x,t)-\dots,
\label{SENO}
\end{eqnarray}
where $f(x,t)=f_s(x,t)+f_c(x,t)$, $f_s(x,t)=\sin(\Omega_J t+
\Gamma x)$, $\eta+\alpha\Omega_2=\alpha\Omega_J$,
\begin{equation}
f_c(x,t)=\frac{1}{2}\sum_{n=0}^{\infty}\left[
A_{1n}\cos(2\Omega_J t+\Gamma x)+
B_{1n}\sin(2\Omega_J t+\Gamma x)\right]\cos(k_n x),
\label{fc}
\end{equation}
\begin{equation}
\alpha\Omega_2=\frac{1}{L}\int\limits_0^{L}
\frac{1}{2}\sum_{n=0}^{\infty} \left[A_{1n}\cos(\Gamma x)-
B_{1n}\sin(\Gamma x)\right]\cos(k_n x)dx,
\label{aw}
\end{equation}
$A_{1n}$ and $B_{1n}$ are:
\begin{equation}
A_{1n}=(2-\delta_{0,n})\frac{\alpha\Omega_J I_C-
({k}_{n}^2-\Omega_J^2)I_S}
{(\alpha\Omega_J)^2+({k}_{n}^2-\Omega_J^2)^2},
\label{An}
\end{equation}
\begin{equation}
B_{1n}=-(2-\delta_{0,n})\frac{\alpha\Omega_J I_S+
({k}_{n}^2-\Omega_J^2)I_C}
{(\alpha\Omega_J)^2+({k}_{n}^2-\Omega_J^2)^2},
\label{Bn}
\end{equation}
and $I_C=\frac{\Gamma L \sin(\Gamma L)\cos(\pi n)}{(\Gamma
L)^2-(\pi n)^2}$, $I_S=\frac{\Gamma L (1-\cos(\Gamma L)\cos(\pi
n))}{(\Gamma L)^2-(\pi n)^2}$, $k_n=\pi n/L$,
$\Omega_J=\eta\omega_J/\alpha$ is the oscillation frequency.

Let us focus on the equation for the second-order frequency
correction (\ref{w2}), that in original variables,
substituting explicit form of $I_C$ and $I_S$, looks like:
\begin{equation}
\Omega_2=-\sum\limits_{n=0}^{\infty}
\frac{(2-\delta_{0,n})(\eta/\alpha+\Omega_2)
(\Gamma L)^2[1-\cos(\Gamma L)\cos(\pi n)]}
{[(\eta+\alpha\Omega_2)^2+
(k_n^2-(\eta/\alpha+\Omega_2)^2)^2][(\Gamma L)^2-(\pi n)^2]^2}.
\label{Om2}
\end{equation}
This transcendental equation may be easily solved and
the voltage-current characteristic ($IVC$) of FFO (due to Josephson
relation the voltage is proportional to $\Omega_J$)
$\Omega_J=\eta/\alpha+\Omega_2$ may be found, see Fig. 1,
where results of computer simulation of Eq. (\ref{PSGE})
and Eq. (\ref{Om2}) are presented for $\alpha=0.2; 0.5$,
$L=5$, $\Gamma=3$.
On the other hand, expressing bias current $\eta$ via
$\Omega_J$ and $\Omega_2$ we arrive to exactly the same
expression for the current-voltage characteristic derived in
\cite{cgssv98},\cite{ss99}:
\begin{equation}
\eta=\alpha\Omega_J+\sum\limits_{n=0}^{\infty}
\frac{(2-\delta_{0,n}) \alpha\Omega_J
(\Gamma L)^2[1-\cos(\Gamma L)\cos(\pi n)]}
{[(\alpha\Omega_J)^2+
(k_n^2-\Omega_J^2)^2][(\Gamma L)^2-(\pi n)^2]^2}.
\label{eta}
\end{equation}
So, if one needs to obtain the function $\Omega_J(\eta)$,
Eq. (\ref{eta}) is more useful. If, however, the function
$\Omega_J(\Gamma)$ (voltage versus control line current) is
of importance, one can use Eq. (\ref{Om2}).
When necessary (e.g., when $\eta/\alpha$ is of the order of
unity), one can recourse to the 4-th and higher order
approximations and derive $IVC$ with the desired precision.

The Eq. (\ref{SENO}) is the equation for slow component of the
phase in the sense that it is considered in the steady-state
limit for $t\to\infty$. Namely such equation is required to
derive different steady-state characteristics, such as
correlation functions and spectra.
Eq. (\ref{SENO}) is linear with respect to
$\psi(x,t)$, but nonlinear with respect to $\eta$ and $\Gamma$.
Substituting now $\eta=\eta_0+\eta_F(x,t)$,
$\Gamma=\Gamma_0+\Gamma_F(x,t)$ (in our model $\Gamma_F(x,
t)=\sigma\eta_F(x,t)$) and
$\psi(x,t)=\psi_0(x,t)+\tilde{\psi}(x,t)$ into (\ref{SENO}) and
linearizing it with respect to small fluctuations ($f(x,t,\eta,
\Gamma)=f(x,t,\eta_0,\Gamma_0)+\left.\frac{\partial f(x,t,\eta,
\Gamma_0)}{\partial \eta}\right|_{\eta=\eta_0}\eta_F(x,t)+
\left.\frac{\partial f(x,t,\eta_0,\Gamma)}
{\partial \Gamma}\right|_{\Gamma=\Gamma_0}\Gamma_F(x,t)+\dots$),
taking into account that $r_d=\left.\frac{\partial
\Omega_J(\eta,\Gamma_0)}
{\partial \eta}\right|_{\eta=\eta_0}$ and $r^{CL}_d=
\left.\frac{\partial \Omega_J(\eta_0,\Gamma)}{\partial
\Gamma}\right|_{\Gamma=\Gamma_0}$, we will get the following
equation for the correction of the phase due to effect of
fluctuations $\tilde{\psi}(x,t)$:
\begin{equation}
\tilde\psi _{tt}+\alpha \tilde\psi _{t_{{}}}-\tilde\psi _{xx}=
(r_d+\sigma r^{CL}_d)\left[\alpha-\frac{\partial f(x,t)}
{\partial \Omega_J} \right]\eta_F(x,t),
\label{SGESN}
\end{equation}
where $r_d$ and $r^{CL}_d$ are dimensionless dynamical
resistances of FFO and control line, respectively.
Following \cite{mal}, the linearization with respect to
small fluctuations can be done if in the area of evolution of the
fluctuating parameter $\eta=\eta_0+\eta_F(x,t)$ bifurcation
points are not located, that corresponds to the previously
assumed condition of the smallness of fluctuations:
fluctuations are so small that do not affect current-voltage
characteristic and, therefore, do not change qualitative behavior
of FFO. Equation (\ref{SGESN}) is more general than its
derivation. In spite it is obtained in the second order
approximation, all quantities in rhs of (\ref{SGESN}) are
fundamental and improving the approximation up to higher order
will only improve quantitative values of $r_d$ and $r^{CL}_d$ and
will add components with $\cos 3\Omega_J t$, $\cos 4\Omega_J t$
and so on into $f(x,t)$. The dynamical resistances are originated
from the assumption of small noise: fluctuations feel the system
as linear if their variance is small at the scale of nonlinearity
and current fluctuations are linearly converted into voltage
fluctuations via transfer factor that is the derivative of the
current-voltage characteristic at the working point
and are not connected with "adiabatic approximation": the noise
in rhs of (\ref{SGESN}) is wideband, but the spectrum of
$\tilde\psi(x,t)$ depends both on properties of
$\eta_F(x,t)$ and the differential operator of the lhs of
(\ref{SGESN}).
If we will consider the short junction limit of (\ref{SGESN})
($\tilde\psi _{xx}=0$, $r^{CL}_d=0$), neglect by parametric
effects $f(x,t)=0$ and change variables: $\tilde v=
\tilde\psi_{t}$, $\tau=\alpha r_d t$, we will get the equation
for fluctuational component of voltage $\tilde v$, presented in
\cite{lik} (Eq. 4.34, page 106).

The solution of Eq. (\ref{SGESN}) can be expressed as a Fourier
serie in space:
$\tilde{\psi}(x,t)=\sum\limits_{m=0}^{\infty}A_m(t)\cos(k_m x)$, where
$k_m=\pi m/L$. Substituting this anzats into (\ref{SGESN}),
multiplying by $\cos(k_n x)$ and integrating
$\frac{1}{L}\int\limits_0^L$, one can get the following equation
for $A_m(t)$:
\begin{equation}
\frac{d^2{A}_{m}}{dt^{2}}+\alpha
\frac{d{A}_{m}}{dt}+k_{m}^{2}{A}_{m} =\xi_m(t),
\label{anoise}
\end{equation}
where $\xi_{m}(t)$ is the projection of the
noise along the $k_{m}$ mode:
\begin{equation}
\xi_{m}(t)=\frac{2-\delta_{0,m}}{L}\int_{0}^{L}
(r_d+\sigma r^{CL}_d)\left[\alpha-\frac{\partial f(x,t)}
{\partial \Omega_J} \right]\eta_F(x,t)\cos (k_{m}x)dx.
\label{xim}
\end{equation}

Let us analyze correlation and spectral properties
of random process $\xi_{m}(t)$. This is nonstationary process
due to periodic time dependence of $f(x,t)$
and its autocorrelation function $\left<\xi_m(t)
\xi_m(t+\tau)\right>$ depends on current time $t$.
Moreover, the product $\frac{\partial f(x,t)}{\partial \Omega_J}
\eta_F(x,t)$ formally is a process linearly growing in time and
does not belong to neither second nor third kind random processes
that complicates the analysis. Here we can use one effective
trick: let us analyze statistical properties of the process
$\zeta_m(t)=\frac{2-\delta_{0,m}}{L}\int_{0}^{L}
(r_d+\sigma r^{CL}_d)f(x,t)\eta_F(x,t)\cos (k_{m}x)dx$
that belongs to the third kind if the
stationary process $\eta_F(x,t)$ is white noise and later
we will take derivative over $\Omega_J$ from the square root
of the intensity of the noise $\zeta_m(t)$. Using this
procedure we can apply the standard technique \cite{mal} and
obtain the correlation function of the second kind of
the process $\zeta_m(t)$:
\begin{equation}
\Phi_{\zeta}(\tau)=\lim_{T^*\to\infty}\frac{1}{2T^*}
\int\limits_{-T^*}^{T^*}
\left<\zeta_m(t)\zeta_m(t+\tau)\right>dt,
\end{equation}
that will lead to stationary delta-correlated process with some
intensity, and finally we get the following correlation function
for the process $\xi_m(\tau)$:
\begin{equation}
\Phi_{\xi m}(\tau)=\alpha^2(r_d+\sigma r^{CL}_d)^2
D_m \delta(\tau),
\end{equation}
where
\begin{equation}\label{dm}
D_{m}=\frac{2k_B T\omega_p}{R_N I_c^2}
\left( 2-\delta _{0,m}\right)\left\{1+H_m \right\},
\end{equation}
\begin{equation}
H_m=\frac{( 2-\delta _{0,m})}{8\alpha^2}
\left[\frac{\partial}{\partial
\Omega_J}\sqrt{\frac{1}{L}\int\limits_0^L\left[\left(
\sum\limits_{i=0}^\infty A_{1n}\cos k_n x \right)^2+
\left(\sum\limits_{i=0}^\infty B_{1n}\cos k_n x \right)^2
 \right](\cos k_m x)^2 dx}\right]^2.
\label{Hm}
\end{equation}
Let us note, that
\begin{equation}\label{D0}
D_{0}=\frac{2k_B T\omega_p}{R_N I_c^2}\left\{1+H_0\right\}
\end{equation}
for $H_0=0$ completely coincides with the dimensionalized noise
intensity for a short (lumped) Josephson junction (see
\cite{lik}). The term $H_m$ comes from down conversion of the
2-nd harmonic due to multiplication of $\cos(2\Omega_J t+\Gamma
x)$ and $\sin(2\Omega_J t+\Gamma x)$ by the noise term
$\eta_F(x,t)$. This effect, as proven in \cite{mal}, takes place
if the power spectral density of the process $\eta_F(x,t)$ is so
wide that significantly different from zero at $2\Omega_J$ and
higher. The term describing the effect of the first harmonic
$\frac{\partial \sin(\Omega_J t+\Gamma x)}{\partial \Omega_J}
\eta_F(x,t)$ will not give any (additional) contribution, since
the intensity of the fluctuational process $\mu(t)= \sin(\Omega_J
t+\Gamma x) \eta_F(x,t)$, $\Phi_{\mu_m}(\tau)=D_m^*\delta(\tau)/2$
will be constant as function of $\Omega_J$ and its derivative
over $\Omega_J$ will give zero. Therefore, we have classical
effect \cite{mal} of the additional parametric broadening of the
main harmonic at the frequency $\omega=\Omega_J$ due to effect of
higher harmonics (at $2\Omega_J$, $3\Omega_J$ and so on).

Since for practical FFOs the second harmonic is
rather weak (the output signal is nearly sinusoidal),
and since, as it will be demonstrated below, noise
components $\xi_m(t)$ with $m\ne 0$ have rather small
effect on fluctuational characteristics of FFO, we
for simplicity of analysis will neglect the term
$H_m$ with $m\ne 0$. For $m=0$ $H_0$ takes the form:
\begin{equation}
H_0=\frac{\left(\eta-\Omega_J/r_d\right)^2}
{16(\alpha\Omega_J)^3(\eta-\alpha\Omega_J)}.
\label{H0}
\end{equation}
In Fig. 2 we present the plot of the ratio between amplitudes of
the second and first harmonics $A_2/A_1$ (diamonds) and the
excess noise $H_0$ (crosses) as functions of $\Omega_J$ using the
approximate Eq. (\ref{Om2}) for $\alpha=0.5$, $L=5$, $\Gamma=3$.
It is seen, that with increase of oscillation frequency
$\Omega_J$ the second harmonic amplitude becomes smaller than the
first one (our approximation works better in high-frequency
limit), but slightly increases at Fiske and Eck steps. The same
qualitative behavior demonstrates the excess noise intensity
$H_0$. Unfortunately, the expression (\ref{H0}), in spite of its
general form, is of restricted usefulness and may give incorrect
results at Fiske steps. This is due to the fact that $H_m$ is
derived in the second order approximation only, but it contains
the factor $1/r_d$. As it is seen from Fig. 1, at Fiske steps the
approximation (\ref{Om2}) gives significantly underestimated
values of dynamical resistance $r_d$ that leads to overestimated
values of $H_0$. Contrary, at the Eck step, where approximation
(\ref{Om2}) and results of computer simulation nearly coincide,
Eq. (\ref{H0}) gives adequate description of excess noise term
$H_0$. In the short junction limit $L\to 0$ one can get the
following approximate expression for the excess noise term:
\begin{equation}
H_{0s}=\frac{\left(\alpha^2+2 \Omega_J^2\right)^2}
{8\alpha^2\Omega_J^4(\alpha^2+\Omega_J^2)^3}.
\label{H0s}
\end{equation}

\section{Form and width of spectral line of FFO}

In this section we will consider the influence of noise
on broadening of spectral line of the FFO.
Following the general setup of the problem \cite{mal},
we will consider the output signal at the end of the junction
in the form:
$v(L,t)=\Omega_J+\nu(t)+R_0\cos(\Omega_J t+\int\nu(t)dt)$,
where $\nu(t)$ are frequency fluctuations,
and $\int\nu(t)dt$ is supposed to be slow process in
comparison with $\cos(\Omega_J t)$ (we do not consider here
amplitude fluctuations, $R_0=const$, since it is known \cite{mal}
that they lead only to some noisy pedestal and do not influence
the linewidth).

The correlation function of frequency fluctuations may be
derived from equation (\ref{anoise}) (see, e.g. \cite{J}).
Since the slow component of the effective noise $\xi_m(t)$ in
(\ref{anoise}) is stationary, for $m\ne 0$ we can write the
equation for autocorrelation function of $A_m(t)$,
$K_{A_{m}}[\tau]=\left<A_{m}(t)A_{m}(t+\tau)\right>$ which
depends only on time difference $\tau$:
\begin{equation}
\frac{d^2 K_{A_{m}}[\tau]}{d\tau^{2}}+\alpha
\frac{dK_{A_{m}}[\tau]}{d\tau}+k_{m}^{2}K_{A_{m}}[\tau]
=0,  \label{kfa}
\end{equation}
and should be solved with the following initial conditions:
\begin{eqnarray}\nonumber
\frac{d^2K_{A_{m}}[\tau]}{d\tau^2}\left|_{\tau=0}\right.=
-\alpha (r_d+\sigma r^{CL}_d)^2 D_{m}/2, \,\, K_{A_{m}}[0]=\alpha
(r_d+\sigma r^{CL}_d)^2 D_{m}/(2k_{m}^{2}).
\end{eqnarray}

The correlation function of $\tilde{\psi}(L,t)$,
$K_{\psi}[\tau]$, may be expressed as follows:
\begin{equation}
K_{\psi}[\tau]=\sum_{m=0}^{\infty}K_{A_{m}}[\tau],
\end{equation}
since $\cos^2(k_m L)=1$.
By the property of the correlation function, the correlation
function of frequency fluctuations
$\nu(t)=\frac{d\tilde{\psi}(L,t)}{dt}$,
$K_{\nu}[\tau]$, is the negative second derivative of
$K_{\psi}[\tau]$:
\begin{eqnarray}\label{knud}
K_{\nu}[\tau]=-\frac{d^2K_{\psi}[\tau]}{d\tau^2}.
\end{eqnarray}

One can see, that for $m=0$ the correlation function
$K_{A_{m}}[\tau]$ diverges, that reflects the only fact, that
for the mode $m=0$ the process $A_0(t)$ is nonstationary. Namely
divergence of $K_{\psi}[\tau]$ (in our case due to divergence of
$K_{A_{0}}[\tau]$) leads to finite linewidths of oscillators
\cite{mal}: in the opposite case when $K_{\psi}[\tau]$ is finite
the linewidth will be zero. The divergence of $K_{A_{0}}[\tau]$
does not lead to any mathematical difficulties since we need to
obtain $\left<\dot{A}_{0}(t)\dot{A}_{0}(t+\tau)\right>$ that is
finite and may be derived from $K_{A_{m}}[\tau]$, using
(\ref{knud}) and limiting transition for $m\to 0$.

Finally, one can get the following expression for the correlation
function of frequency fluctuations:
\begin{eqnarray}  \label{knu}
\begin{array}{cc}
K_{\nu}[\tau]=\alpha (r_d+\sigma r^{CL}_d)^2 D_{0}\times \\
\left\{\frac{1}{2} e^{-\alpha\tau} + e^{-\alpha\tau/2}
\sum\limits_{m=1}^{\infty} \left[\cos(f(\alpha,k_m)\tau)-
\frac{\alpha}{2f(\alpha,k_m)} \sin(f(\alpha,k_m)\tau)
\right]\right\},
\end{array}
\end{eqnarray}
where
$f(\alpha,k_m)=f(k_m)=\sqrt{k_m^2-\left(\frac{\alpha}{2}\right)^2}$.
If $k_m^2<\left(\frac{\alpha}{2}\right)^2$, then one has to change
$\sin$ and $\cos$ to the corresponding hyperbolic functions.

For stationary and Gaussian frequency fluctuations
(since Eq. (\ref{anoise}) is linear with Gaussian noise,
the probability distribution of $A_m$ is also Gaussian),
the form of spectral line may be written as \cite{mal}:
\begin{equation}
W_v(\omega)=\frac{R_0^2}{4\pi}
\int\limits_{-\infty}^{+\infty}\exp\left[-\chi(\tau)\right]
\cos\omega\tau d\tau,
\label{spf}
\end{equation}
where $\chi(\tau)$ is statistical structural function that is
nonnegative and even function of $\tau$:
\begin{equation}
\chi(\tau)=\frac{1}{2}\int\limits_{-\tau}^{+\tau}
(\tau-|\xi|)K_{\nu}[\xi]d\xi.
\label{chi}
\end{equation}
Substituting $K_{\nu}[\tau]$ (\ref{knu}) into (\ref{chi}), we get:
\begin{eqnarray} \label{chi2}
\begin{array}{cc}
\chi(\tau)=\frac{1}{2}(r_d+\sigma r^{CL}_d)^2 D_{0}\times \\
\left\{ \tau + \frac{e^{-\alpha\tau}-1}{\alpha} +
\frac{\alpha L^2}{3}- e^{-\alpha\tau/2}\sum\limits_{m=1}^{\infty}
\frac{2\alpha}{k_m^2}\left[\cos(f(k_m)\tau)+ \frac{\alpha
\sin(f(k_m)\tau)}{2f(k_m)}
 \right] \right\}, \,\, \tau\ge 0.
\end{array}
\end{eqnarray}
One can check, that the statistical structural function
(\ref{chi2}) that we will use for calculation of the linewidth
and the form of spectral line is a smooth finite function
and the sum in (\ref{chi2}) is converging due to the term
$1/k_m^2$.

For the case of stationary Gaussian fluctuations of frequency
the linewidth is defined in the following way \cite{mal}:
\begin{equation}
\Delta\Omega=\frac{\pi}{\int\limits_0^\infty \exp[-\chi(\tau)]d\tau}.
\label{lwd}
\end{equation}
Substituting the structural function (\ref{chi2}) into (\ref{lwd}),
we get the following final expression for the linewidth:
\begin{equation}
\Delta\Omega=\frac{\pi}
{\int\limits_0^\infty \exp[-F_1(\tau)-
F_2(\tau)]d\tau}, \label{lw}
\end{equation}
where
\begin{equation}
F_1(\tau)=\frac{1}{2}(r_d+\sigma r^{CL}_d)^2 D_{0}
\left[ \tau + \frac{e^{-\alpha\tau}-1}{\alpha}\right],
\label{lw1}
\end{equation}
\begin{equation}
F_2(\tau)=\frac{1}{2}(r_d+\sigma r^{CL}_d)^2 D_{0}
\left[\alpha \frac{L^2}{3}-
e^{-\alpha\tau/2}\sum\limits_{m=1}^{\infty}
\frac{2\alpha}{k_m^2}\left[\cos(f(k_m)\tau)+ \frac{\alpha
\sin(f(k_m)\tau)}{2f(k_m)}
 \right]\right]. \label{lw2}
\end{equation}

In the following we will not recourse to two known limiting cases
of very fast and very slow frequency fluctuations (Lorentzian and
Gaussian form of the spectral line) but perform the exact
analysis of linewidth and spectral form on the basis of
expressions (\ref{spf})-(\ref{lw2}).

Let us analyze the linewidth given by expressions
(\ref{lw})-(\ref{lw2}). First, consider the case of a short
Josephson junction $L\to 0$, $r^{CL}_d=0$, neglecting by
parametric effects $H_m=0$. Then the function $F_2(\tau)$
disappears: $F_2(\tau)=0$. If damping coefficient $\alpha$ is
large (but, of course, $\eta/\alpha > 1$), one can neglect the
term $(e^{-\alpha\tau}-1)/\alpha$ in (\ref{lw}),(\ref{lw1}) and
then well-known expression for the linewidth in Lorentzian
approximation may be obtained:
\begin{equation}
\Delta\Omega_s=\frac{\pi}{2}r_d^2 D_{0},
\label{lwa}
\end{equation}
that in dimensional units looks like:
\begin{equation}
\Delta f_s=\frac{1}{2}\left(\frac{2\pi}{\Phi_0}\right)^2
R_d^2 \frac{k_B T}{R_N}. \label{lwad}
\end{equation}
It should be noted, that this expression is larger than one in
\cite{RS} by a factor of $\pi/2$ (see formula (12) in \cite{RS},
we neglected by $I_p$ and $R_N=V_0/I_{qp}$ in their notations).
This may be explained by different definitions of the linewidth:
we define it as the width of rectangle with the equal square
\cite{mal}, that for Lorentzian form of spectral line should give
just by a factor of $\pi/2$ larger value than the definition of
the linewidth at $1/2$ level, that was used in \cite{RS}. In the
following we will present all plots of the linewidth
(\ref{lw})-(\ref{lw2}) multiplied by the factor $2/\pi$ for
correct comparison with experiment where the linewidth is defined
at the level $1/2$.

If, however, damping coefficient $\alpha$ is rather small, there
may be significant deviation of the linewidth of a short junction
from (\ref{lwa}) depending on values of $D_0$ and $r_d$ (at given
$D_0$ and $r_d$ the linewidth will be smaller), and the use of
exact formula (\ref{lw}) for $F_2=0$ is necessary.

Now let us consider plots of the exact expression of FFO
linewidth (\ref{lw})-(\ref{lw2}) neglecting by magnetic field
fluctuations $r^{CL}_d=0$. Our aim here is to understand how
spatial modes with $m\ne 0$ influences the linewidth, i.e. how
long junction differs from the short one. The plots of the
linewidth versus dynamical resistance are presented in Fig. 3 in
dimensional units ($MHz$ vs $Ohm$) for practical FFO parameters
\cite{kosh2}: $L=76.24$, $\alpha=0.0074$ ($\alpha L<1$),
$\alpha=0.04$ ($\alpha L>1$), $T=4.2K$, $R_N=0.04 \,Ohm$, the
dynamical resistance is: $R_d=0.01-10 \,Ohm$; we neglected by the
excess noise term $H_0=0$. From Fig. 3 one
can see, that below certain threshold curves for $\alpha L<1$ and
$\alpha L>1$ have the same behavior and coincide with the short
junction case (\ref{lwad}). Above the threshold that depends on
noise intensity (\ref{D0}) one can see the effect of spatial
modes: the curves split and the linewidth for $\alpha L>1$ is
greater than for $\alpha L<1$. With increase of noise intensity
the threshold region located for $T=4.2 K$ between $R_d=0.1-1
\,Ohm$ will move to the left. However, it should be noted that
practical range of dynamical resistance lies from $0.001$ to
$0.1$ $Ohm$, where the effect of spatial modes can be neglected
that confirms our previous assumption to neglect the excess noise
term $H_m$ for $m\ne 0$.

Therefore, it seems that in the practical range of parameters
the linewidth of FFO may be well described by the following
approximate expression that may be derived from
(\ref{lw})-(\ref{lw2}) neglecting by spatial modes with
$m\ne 0$ and by the term $(e^{-\alpha\tau}-1)/\alpha$:
\begin{equation}
\Delta f_{FFO}=\frac{1}{2}\left(\frac{2\pi}{\Phi_0}\right)^2
(R_d+\sigma R^{CL}_d)^2 \frac{k_B T}{R_N} (1+H_0), \label{lwffo}
\end{equation}
where $H_0$ is given by (\ref{H0}). Now let us compare the
expression (\ref{lwffo}) (multiplied by $2/\pi$) with the
experimental results \cite{kosh2}, see Fig. 4. We again take for
simplicity $H_0=0$, as before
$L=76.24$, $T=4.2 K$, $R_N=0.04 \,Ohm$; for $\alpha L<1$:
$\alpha=0.0074$, for $\alpha L>1$:
$\alpha=0.04$. In \cite{kosh2} $R^{CL}_d$ has not been explicitly
measured and we have used $\sigma R^{CL}_d$ as fitting parameter:
putting the noise conversion factor $\sigma=1$ we have chosen
$R^{CL}_d$ to fit experimental results only at one point for
$R_d\to 0$: for $\alpha L<1$ $R^{CL}_d=0.006 \,Ohm$ and
for $\alpha L>1$ $R^{CL}_d=0.04 \,Ohm$. One can see good
agreement between expression (\ref{lwffo}) and the results of
experiment. More detailed comparison of expressions
(\ref{lw})-(\ref{lw2}) and (\ref{lwffo}) with experimental
results obtained in different layouts, taking into account the
excess noise term $H_0$ will be given elsewhere.

In fact, Fig. 4 and the expression (\ref{lwffo}) give an idea how
the noise conversion factor $\sigma$ may be measured. If noise
intensity is known, setting $R_d\ll R_d^{CL}$ one can get the
value $(\sigma R_d^{CL})^2$ from experimentally measured plot
of the linewidth. On the other hand, the value $R_d^{CL}$ is
independently accessible from experiment.

Let us consider the form of spectral line of the FFO.
In general \cite{mal}, the spectral line consists of narrow and high
spectral peak that finite width is originated by frequency
fluctuations (nonstationary phase fluctuations) and broad and low
pedestal due to amplitude fluctuations. If amplitude and
frequency fluctuations are correlated but small, there will be
also small asymmetric contributions both into the peak and the
pedestal. Since in the frame of the present paper we do not
consider amplitude fluctuations, below we will consider the form
of spectral peak, that may be derived from (\ref{spf}),
substituting statistical structural function (\ref{chi2}).

Since in practically interesting range of parameters the
linewidth is well described by formula (\ref{lwffo}), we will
also neglect by spatial modes $m\ne 0$ in (\ref{chi2}) and by
$(e^{-\alpha\tau}-1)/\alpha$ and will get well-known expression
for the Lorentzian form of spectral line:
\begin{equation}
W_v(\omega)=\frac{R_0^2}{4\pi}\frac{2(\Delta f_{FFO}/\pi)}
{(\Delta f_{FFO}/\pi)^2+\omega^2},
\label{lor}
\end{equation}
the width of this curve at the level $1/2$ is given by
$2(\Delta f_{FFO}/\pi)$, where $\Delta f_{FFO}$ is given by
(\ref{lwffo}). In Fig. 5 plots of spectral form, given by
(\ref{spf}),(\ref{chi2}) (solid lines) and the approximation
(\ref{lor}) (dashed lines) are presented. It is seen, that for a
small dynamical resistance ($R_d=0.001$ $Ohm$, that corresponds to
plato in Fig. 4) the curves absolutely coincide. With increase of
$R_d$ the exact expression slightly deviates from the Lorentzian
approximation, but in all practical range of parameters (up to
$R_d=0.1$ $Ohm$), formula (\ref{lor}) gives adequate description of
the form of spectral line. For larger $R_d$ and noise intensity,
the deviation from Lorentzian form will increase and it is
necessary to use expressions (\ref{spf}),(\ref{chi2}). Note, that
in \cite{bk97} Lorentzian form of spectral line of FFO was
predicted using another approach. Recently, Lorentzian form of
spectral line of FFO has been experimentally observed in wide
range of parameters both at Fiske and Eck steps \cite{koshisec}.

If we would separately consider the case of technical
fluctuations (since these fluctuations are slow and narrowband
the consideration may be performed in the adiabatic
approximation and in this case the effect of additional
parametric broadening of the linewidth would not appear due to
the fact that the spectrum of technical fluctuations is much
more narrow than the basic frequency of the FFO), the form of the
spectral line will be Gaussian (for Gaussian distributed
technical fluctuations) and the linewidth will be given as
$\sqrt{2\pi \left<\nu^2\right>}$, where $\left<\nu^2\right>$ is
variance of frequency fluctuations. It should be noted, that if
one wishes to consider the joint effect of natural and technical
fluctuations (as they coexist in real life), one can not recourse
to the Lorentzian and Gaussian limiting cases, but the
calculation of the spectral form should be performed using
formula (\ref{spf}).

\section{Conclusions}

The aim of the present paper is to investigate the influence
of wideband fluctuations of bias current and magnetic field on
dynamics of FFO. We have derived analytical
expressions both for the form of spectral line of FFO and its
width. For practical range of parameters simple approximate
expression of the linewidth, that well fits the experimental
results, has been given. The appearance of excess noise (in
comparison with the short junction linewidth) has been explained
by presence of magnetic field fluctuations and by the so-called
parametric broadening of spectral line due to influence of higher
harmonics. It has been demonstrated that in the practical range
of parameters, in the case when thermal fluctuations dominate,
the Lorentzian form of spectral line is realized, while for
larger values of dynamical resistance and temperature deviations
from Lorentzian form may be observed.

\section{Acknowledgments}

The author wishes to thank V. P. Koshelets, V. V. Kurin, N. Likhanov,
A. Lukyanov, A.~N.~Malakhov, J. Mygind, M. Salerno, M. Samuelsen and
A. Yulin for helpful discussions and
the Dpt. of Physics "E.R.Caianello" of the university of Salerno
and personally Prof. Mario Salerno for the offered position where
significant part of this work has been done.
The work has been supported by the Russian Foundation for Basic
Research (Project N~99-02-17544, Project N~00-02-16528 and
Project N~00-15-96620), by INFM (Istituto Nazionale di Fisica
della Materia) and by the MURST (Ministero dell'Universita' e
della Ricerca Scientifica e Tecnologica).


\newpage
\vspace*{\fill}
\begin{figure}[th]
\centerline{
\epsfxsize=12cm
\epsffile{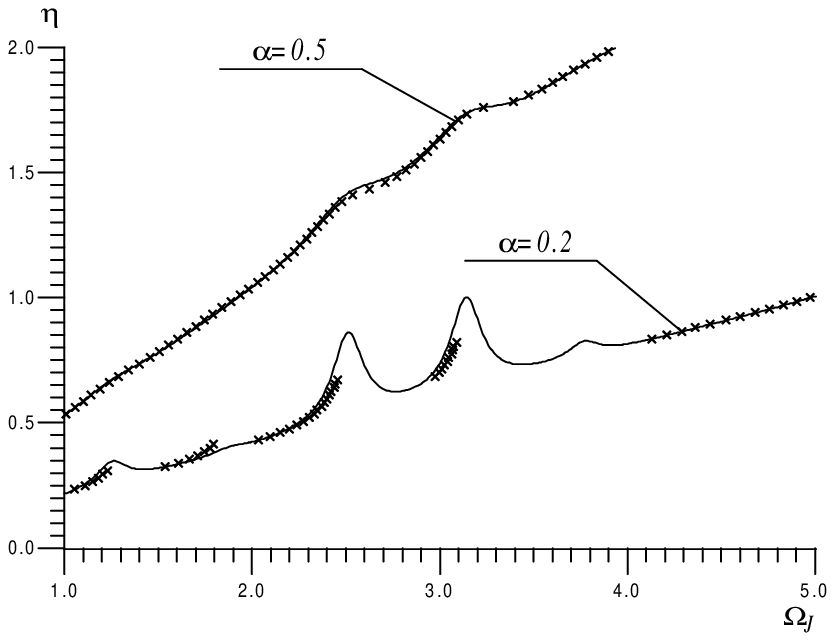}}
\vspace{5pt}
\caption[b]{\label{fig1}
Current-voltage characteristic. Numerical solution
of the sine-Gordon equation is presented by crosses
and the second order approximation is given by solid
line for $L=5$, $\Gamma=3$, $\alpha=0.2; 0.5$.
}
\end{figure}
\vspace*{\fill}

\newpage
\vspace*{\fill}
\begin{figure}[th]
\centerline{
\epsfxsize=12cm
\epsffile{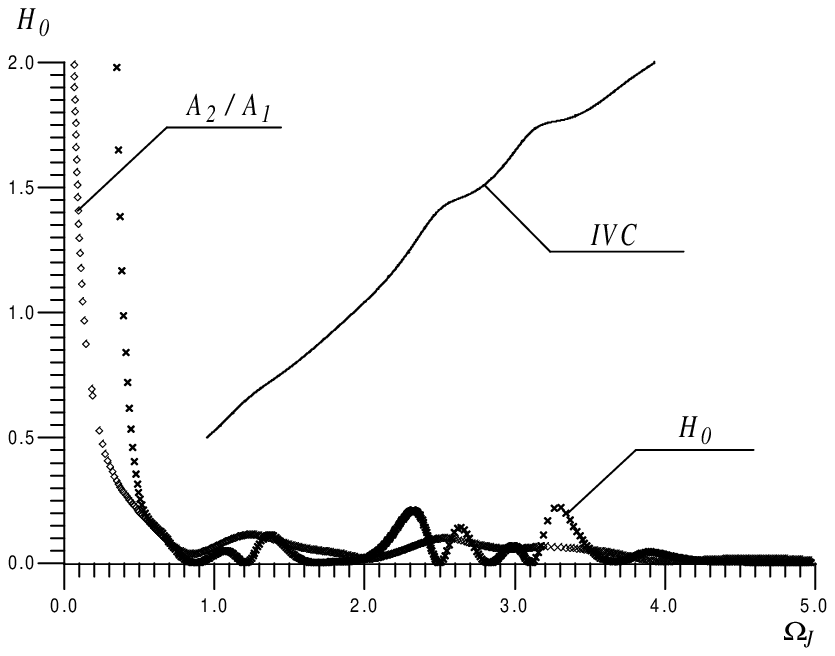}}
\vspace{5pt}
\caption[b]{\label{fig2}
The excess noise intensity $H_0$ (crosses) and the
ratio between amplitudes of the second and first
harmonics $A_2/A_1$ (diamonds) for $L=5$, $\Gamma=3$,
$\alpha=0.5$. For comparison the corresponding
current-voltage characteristic ($IVC$) is given (solid line).
}
\end{figure}
\vspace*{\fill}

\newpage
\vspace*{\fill}
\begin{figure}[th]
\centerline{
\epsfxsize=12cm
\epsffile{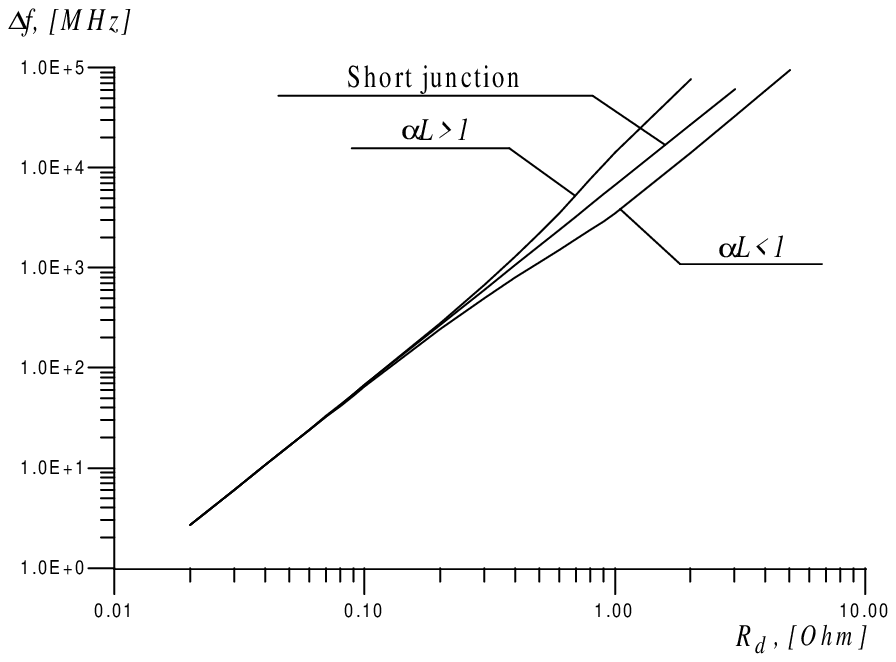}}
\vspace{5pt}
\caption[b]{\label{fig3}
Illustration of the effect of splitting of the linewidth
for $\alpha L <1$ and $\alpha L >1$.
}
\end{figure}
\vspace*{\fill}

\newpage
\vspace*{\fill}
\begin{figure}[th]
\centerline{
\epsfxsize=12cm
\epsffile{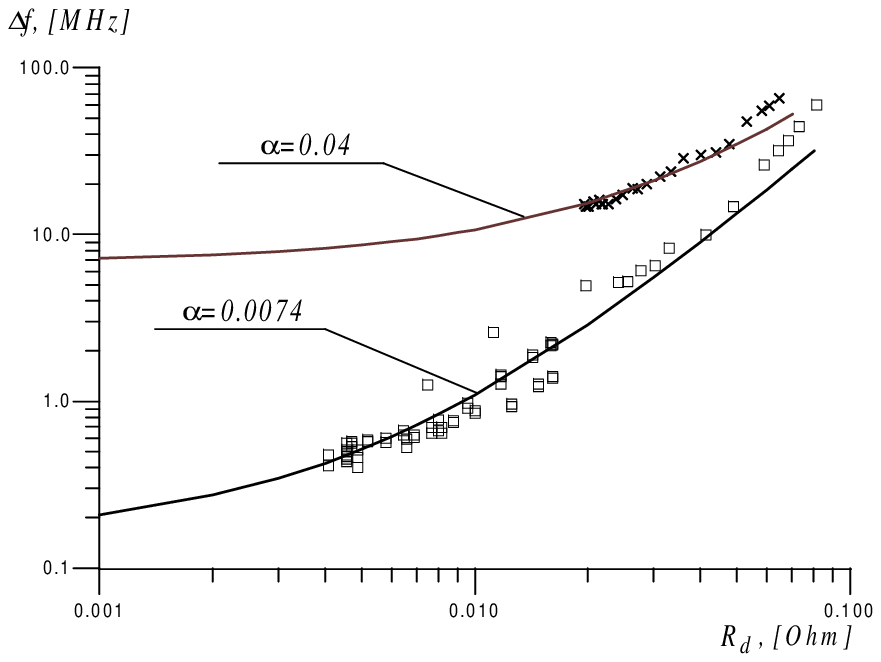}}
\vspace{5pt}
\caption[b]{\label{fig4}
Comparison of experimental and theoretical linewidths for the
parameters: $L=76.24$, $T=4.2 K$, $R_N=0.04 \,Ohm$, $\sigma=1$;
for $\alpha L<1$: $\alpha=0.0074$, $R^{CL}_d=0.006 \,Ohm$; for
$\alpha L>1$: $\alpha=0.04$, $R^{CL}_d=0.04 \,Ohm$;
solid lines - theory, crosses and squares - experimental results.
}
\end{figure}
\vspace*{\fill}

\newpage
\vspace*{\fill}
\begin{figure}[th]
\centerline{
\epsfxsize=12cm
\epsffile{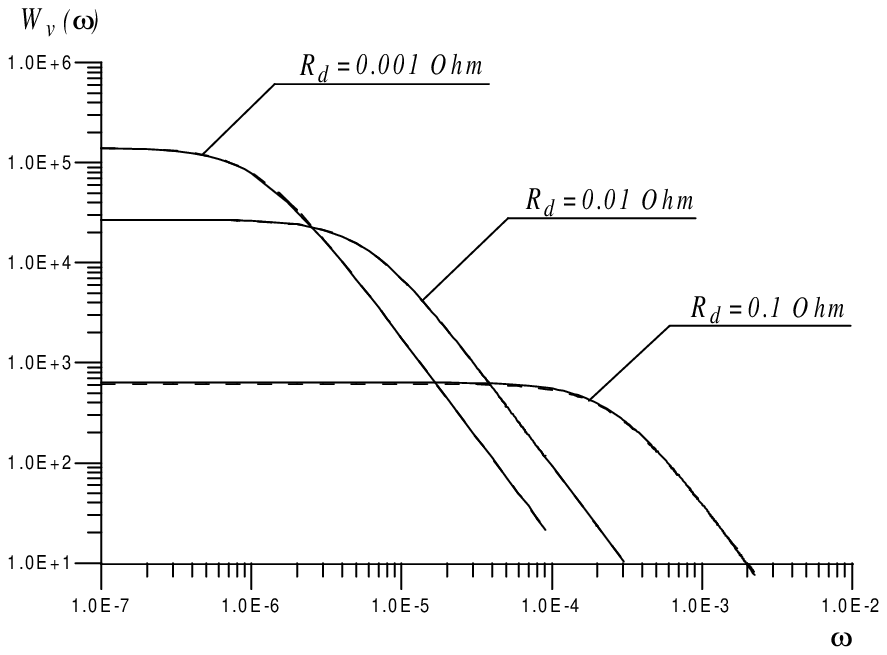}}
\vspace{5pt}
\caption[b]{\label{fig5}
The form of spectral line: comparison of Lorentzian approximation
(dashed lines) and exact expression for the spectral form (solid
lines). It is seen, that the corresponding curves coincide.
}
\end{figure}
\vspace*{\fill}

\end{document}